\documentclass[paper]{JHEP3}
\usepackage{amsmath,amssymb,amsthm,amscd,graphicx}
\usepackage{psfrag}

\newcommand{\CC}{{\cal C}}

\newcommand{\CM}{{\cal M}}
\newcommand{\CN}{{\cal N}}
\newcommand{\CO}{{\cal O}}

\newcommand{\CW}{{\cal W}}

\def\IR{{\mathbb R}}

\def\IP{{\mathbb P}}
\def\IS{{\mathbb S}}

\def\IN{{\mathbb N}}

\newcommand{\tr}{{\rm Tr}}
\newcommand{\re}{{\rm e}}
\newcommand{\ri}{{\rm i}}
\newcommand{\rd}{{\rm d}}

\newcommand{\nn}{\nonumber}

\newcommand{\ssq}[1]{{\sqrt{\sigma({#1})}}}

\newcommand{\bra}{\left\langle}
\newcommand{\ket}{\right\rangle}
\renewcommand{\l}{\left}
\renewcommand{\r}{\right}
\newcommand{\de}{\partial}
\newcommand{\bit}{\begin{itemize}}
\newcommand{\eit}{\end{itemize}}
\newcommand{\al}{\alpha}

\newcommand{\be}{\begin{equation}}
\newcommand{\ee}{\end{equation}}
\newcommand{\ba}{\begin{aligned}}
\newcommand{\ea}{\end{aligned}}

\newcommand{\sectiono}[1]{\section{#1}\setcounter{equation}{0}}

\newcommand{\bea}{\begin{eqnarray}}
\newcommand{\eea}{\end{eqnarray}}
\newcommand{\bean}{\begin{eqnarray*}}
\newcommand{\eean}{\end{eqnarray*}}
\newcommand{\bdm}{\begin{displaymath}}
\newcommand{\edm}{\end{displaymath}}

\newcommand{\ben}{\begin{eqnarray}\displaystyle}
\newcommand{\een}{\end{eqnarray}}

\newtheorem{theorem}{Theorem}[section]

\theoremstyle{definition}

\newtheorem{example}[theorem]{Example}

\newdimen\tableauside\tableauside=1.0ex
\newdimen\tableaurule\tableaurule=0.4pt
\newdimen\tableaustep
\def\phantomhrule#1{\hbox{\vbox to0pt{\hrule height\tableaurule width#1\vss}}}
\def\phantomvrule#1{\vbox{\hbox to0pt{\vrule width\tableaurule height#1\hss}}}
\def\sqr{\vbox{%
  \phantomhrule\tableaustep
  \hbox{\phantomvrule\tableaustep\kern\tableaustep\phantomvrule\tableaustep}%
  \hbox{\vbox{\phantomhrule\tableauside}\kern-\tableaurule}}}
\def\squares#1{\hbox{\count0=#1\loop\sqr
  \advance\count0 by-1 \ifnum\count0>0\repeat}}
\def\tableau#1{\vcenter{\offinterlineskip
  \tableaustep=\tableauside\advance\tableaustep by-\tableaurule
  \kern\normallineskip\hbox
    {\kern\normallineskip\vbox
      {\gettableau#1 0 }%
     \kern\normallineskip\kern\tableaurule}%
  \kern\normallineskip\kern\tableaurule}}
\def\gettableau#1{\ifnum#1=0\let\next=\null\else
\squares{#1}\let\next=\gettableau\fi\next}

\newcommand{\figref}[1]{Fig.~\protect\ref{#1}}
\title{The uses of the refined matrix model recursion}

\author{
Andrea Brini, Marcos Mari\~no and S\'ebastien Stevan\\
D\'epartement de Physique Th\'eorique et Section de Math\'ematiques,\\
Universit\'e de Gen\`eve, Gen\`eve, CH-1211 Switzerland
\\
}

\abstract{
We study matrix models in the $\beta$-ensemble by building on the refined recursion
relation proposed by Chekhov and Eynard. 
We present explicit results for the 
first $\beta$-deformed corrections in the one-cut and the two-cut cases, as well as two applications to supersymmetric gauge theories: 
the calculation of superpotentials in $\CN=1$ 
gauge theories, and the calculation of vevs of surface operators in superconformal $\CN=2$ theories and their Liouville duals. Finally, we study the $\beta$-deformation of the Chern--Simons matrix model. Our results indicate that this model does not provide an appropriate description of the $\Omega$-deformed topological string on the resolved conifold, and therefore that the $\beta$-deformation might provide a different generalization of topological string theory in toric Calabi--Yau backgrounds.}    

\begin{document}

\sectiono{Introduction}

Matrix models in the $1/N$ expansion have become a powerful tool in the study of supersymmetric gauge theories and string theories. For example, as shown by Dijkgraaf and Vafa in \cite{dv}, the all-genus free energies of type B topological string theories on certain non-compact Calabi--Yau manifolds can be computed from the $1/N$ expansion of simple, polynomial matrix models, and this leads to exact results for the superpotentials of a large class of $\CN=1$ supersymmetric theories \cite{dvp}. Other applications include the matrix model formulation of Chern--Simons theories \cite{mm,mmhouches} and the matrix model-inspired remodeling of the B-model \cite{mmopen, bkmp} for the mirrors of general toric geometries. As a consequence of these relationships, the recent progress in solving the $1/N$ expansion of matrix models \cite{eynard, eo} has found many applications in string theory and gauge theory.  

Most of these applications involve the standard Hermitian matrix model ensemble. There is a well-known one-parameter deformation of this ensemble, usually called the 
$\beta$-ensemble or the $\beta$-deformation, which involves an extra parameter $\beta$. The standard Hermitian ensemble is obtained when $\beta=1$, and the special values $\beta=2$ and $\beta=1/2$ correspond to ${\rm Sp}(N)$ real quaternionic and ${\rm SO}(N)$ real symmetric matrices, respectively. The $1/N$ expansion for the more general, $\beta$-deformed ensemble, has been worked out in an algebro-geometric language by Chekhov and Eynard \cite{ce,chekhov}. As in \cite{eynard,eo}, explicit expressions for the expansion of correlators and free energies are obtained through a ``refined" recursion relation based on the spectral curve of the matrix model with $\beta=1$\footnote{This recursion has been reformulated in \cite{em,cem} in terms of ``quantum algebraic curves," but the original 
formulation in \cite{ce} is more useful for the purposes of this paper.}. 

The general $\beta$ ensemble also has many applications. For example, the special values $\beta=2,1/2$ lead to the enumeration of non-orientable surfaces (see for example \cite{bn,mw}), and this can be used to construct non-critical unoriented strings in an appropriate double-scaling limit \cite{bn,hm} (see \cite{dfgzj,nakayama} for a review of these ideas). These ensembles also appear naturally when one applies the techniques pioneered by Dijkgraaf and Vafa to 
supersymmetric gauge theories with ${\rm SO}(N)$ and ${\rm Sp}(N)$ gauge symmetry \cite{ks1,ks2,ac}. More recently, there has been renewed interest in the $\beta$-ensemble in the context of the so-called AGT correspondence between 
$\CN=2$ gauge theories and Liouville theory \cite{agt}. In this correspondence, conformal blocks in Liouville theory are identified with $\Omega$-deformed partition functions \cite{nekrasov} in $\CN=2$ theories, and it has been argued in \cite{dvtoda} that the general $\Omega$-deformation of $\CN=2$ superconformal field theories can be implemented by a $\beta$-deformed matrix model with a Penner-type potential. 

In this paper we analyze in detail the recursive proposal by Chekhov and Eynard
and its concrete implementation in various examples. In section 2 we
thoroughly study
the algebro-geometric solution of the loop equations for the $\beta$-deformed
eigenvalue model; in doing so, we first find a correction to the diagrammatic solution
of \cite{ce}, which was also very recently pointed out by Chekhov
\cite{chekhov}, and discuss various technical issues associated to the
$\beta$-deformation with respect to the
ordinary $\beta=1$ case. We moreover present explicit formulae for the very first corrections to correlators and free energies in the $\beta$-ensemble for a variety of situations and potentials; in the one-cut case and for polynomial potentials, some of these formulae were already derived in \cite{gm} and used there to analyze the universality 
properties of the asymptotic enumeration of graphs in non-orientable surfaces (see also the recent paper \cite{eynardformulae} for another derivation of explicit 
one-cut formulae). In section 3 we use these results to study applications to supersymmetric gauge theories. The first application is the computation of $\CN=1$ superpotentials, where we recover and generalize previous results in \cite{ks1,ks2,ac,shigemori2}. 
Our second application is to the AGT correspondence, where we consider surface operators \cite{aggtv,saraetal,dgh} 
in a very simple example associated to a sphere with three punctures. In this case, 
we generalize the B-model computation in \cite{saraetal} and show that the $\beta$-deformed correlators obtained with the formalism of \cite{ce} 
lead to correlation functions in Liouville theory for general background charge. 

One motivation for the present work was to find a matrix model formulation of topological string theory in an 
$\Omega$-background. This background 
provides a one-parameter 
deformation of topological string theory (at least on certain toric
Calabi--Yau manifolds) which was originally obtained via a five-dimensional
version of Nekrasov's partition function \cite{nekrasov}. The $\Omega$-deformed topological string was reformulated later 
on in terms of the refined topological vertex \cite{ikv}. More recently, 
the holomorphic anomaly equation has been generalized to the
$\Omega$-background for $\CN=2$ gauge theories \cite{kw} and more generally
for the A-model on local Calabi Yau manifolds \cite{hk}, 
thus providing an important step towards a B-model version of this deformed theory. 

It is natural 
to try to extend the remodeling of the B-model \cite{bkmp} to this
deformation, and the refined recursion relation of Chekhov and Eynard is a
natural candidate for this, as suggested by the arguments of \cite{dvtoda} and
by our computations in Section \ref{sec:AGT}. 
In order to test this idea we analyze, in section 4, 
the $\beta$-deformed Chern--Simons (CS) matrix model of \cite{mm}. When $\beta=1$ this model is dual 
to type A topological string theory on the resolved conifold, and its
$\beta$-deformation is a natural candidate for the $\Omega$-deformation of
this theory. Our explicit computations, verified by perturbative calculations,
show that the recursion of \cite{ce} works perfectly well for the CS matrix
model\footnote{This is not entirely guaranteed {\it a priori}, as the
  formalism of \cite{ce, chekhov} applies in principle to polynomial or
  at most logarithmic potentials.}, but unfortunately they do not seem compatible with the
$\Omega$-deformation, at least when taken at face value (this was mentioned as
well in \cite{hk}). An interesting feature we discover 
is a highly involved analytic dependence of $\beta$-deformed amplitudes on
the closed string moduli with respect to their ``refined'' counterpart. This
degree of sophistication only increases when moving to multi-cut models, where
the exact formulae we find, {\it e.g.} for the cubic matrix model, display a
more intricated analytic structure as compared to oriented, open
amplitudes at higher genus \cite{eo, bkmp}. In particular, they cannot be
immediately related to the same type of holomorphic quasi-modular forms of the ordinary topological
string in a self-dual background \cite{abk}, and it would be interesting to see
what kind of generalization would be needed to encompass this more general case.

Our work indicates that the matrix model $\beta$-deformation can be defined and computed for the mirrors of 
other toric Calabi--Yau manifolds. An important example are the mirrors of $A_p$ fibrations over $\IP^1$. These models can be described 
by Chern--Simons matrix models on lens spaces \cite{akmv}, and one can generalize the computation performed in section 4 to this more general setting. In fact, it is likely, in view of 
the progress in formulating the $\beta$-deformation in a geometric language \cite{cem}, that the $\beta$-deformation provides a generalization of the B-model for the mirrors 
of toric Calabi--Yaus. According to our explicit results, it seems that this deformation will be in general {\it different} from the $\Omega$-deformation. If this is indeed the case it would be 
interesting to understand more aspects of this deformation. For example, one could use the Chekhov--Eynard recursion, together with the strategy of \cite{emo}, to formulate a 
holomorphic anomaly equation for the $\beta$-deformed free energies. More generally, one should try to understand this deformation in the language of the A-model and in the gauge theory 
language.

\sectiono{Beta ensemble and topological recursion}

In this section we review and analyze the formalism of Chekhov and Eynard \cite{ce}, which proposes a topological recursion for the beta ensemble of random matrices. 
We will discuss the implementation of their formulae and present explicit expressions for various models. 

\subsection{General aspects}
In terms of eigenvalues, the beta ensemble of random matrices is defined by the partition function 
\be
\label{betaz}
Z={1\over  N! (2 \pi)^{N}}  \int \prod_{i=1}^N \rd\lambda_i |\Delta (\lambda)|^{2\beta} \re^{-{\beta \over  g_s} \sum_{i=1}^NV( \lambda_i)}.
\ee
In what follows we will mainly follow the normalizations in \cite{ce}. The connected correlators are defined through 
\be\label{scor}
W (p_1, \ldots, p_h) =g_s^{2-h} \beta^{h-1} \left\langle \tr\, {1\over p_1-M} \cdots \tr\, {1\over p_h-M} \right\rangle^{(\mathrm{c})}, \qquad h\ge 1. 
\ee
The correlator $W(p)$ for $h=1$ is usually called the {\it resolvent} of the matrix model. 
Both the free energies and the connected correlators have an asymptotic expansion in $g_s$, in which the 't Hooft parameters are kept fixed. 
In the case of the free energy $F=\log Z$, we have
\be
\label{genusF}
F=\sum_{k,l\ge 0} g_s^{2k+l-2} \beta^{1-l/2-k} \gamma^l F_{k,l}.
\ee
where
\be
\gamma={\sqrt{\beta}} -{\sqrt{\beta^{-1}}}. 
\ee
For the first few terms we find, explicitly,  
\be
\ba
F&=g_s^{-2} \beta F_{0,0} + g_s^{-1} (\beta-1) F_{0,1} + (\beta+ \beta^{-1} -2) F_{0,2} + F_{1,0} \\ 
&+g_s \left( {( \beta-1)^3\over \beta^2 } F_{0,3} + (1-\beta^{-1}) F_{1,1} \right) \\
&+g_s^2\left( \beta^{-1} F_{2,0}+ {(\beta-1)^2 \over \beta^2} F_{1,2} + {(\beta-1)^4 \over \beta^3} F_{0,4} \right)+\cdots
\ea
\ee
The $g_s$ expansion of the connected correlators is written as 
\be
\label{genusW}
W (p_1, \ldots, p_h) =  \sum_{g=0}^{\infty} \hbar^{2g} W_{g} (p_1, \ldots, p_h),
\ee
where $g$ can be an integer or a half-integer, and $\hbar$ is defined as
\be
\hbar ={g_s\over {\sqrt{\beta}}}.
\ee
This expansion defines the ``genus" $g$ correlators, which can be in turn expanded as 
\be
\label{ggamma}
W_{g} (p_1, \ldots, p_h)=\sum_{k=0}^{[g]} \gamma^{2g-2k} W_{k,2g-2k}(p_1, \cdots, p_h), 
\ee
and leads to the following expansion for connected correlators, 
\be
\langle \tr M^{n_1} \cdots \tr M^{n_h}\rangle^{(c)}=\left( {g_s \over \beta}\right)^h \sum_{g\ge 0} \sum_{k=0}^{[g]} \hbar^{2g-2} \gamma^{2g-2k} \langle \tr M^{n_1} \cdots \tr M^{n_h}\rangle^{(c)}_{k,l}. 
\ee

The beta ensemble might be regarded as a natural deformation of the standard Hermitian ensemble, since when $\beta=1$ (\ref{betaz}) becomes the 
standard partition function of the (gauged) Hermitian matrix model. In this case, in the expansion of the free energy and the correlators only the terms 
$F_{g,0}$ and $W_{g,0}$ contribute (with $g$ a non-negative integer). This leads to the standard expansion in powers of $g_s^2$ of the 
Hermitian matrix model. On the other hand, there are two special values of $\beta$ which have a matrix model realization: for $\beta=1/2$, (\ref{betaz}) describes an ensemble of real symmetric matrices with orthogonal $SO(N)$ symmetry, while the case $\beta=2$ describes an ensemble of quaternionic real matrices with symplectic ${\rm Sp}(N)$ symmetry. 

\begin{example} {\it The Gaussian $\beta$ ensemble}. In the Gaussian case 
\be
V(x)=x^2
\ee
the matrix integral (\ref{betaz}) can be computed at finite $N$ by using Mehta's formula 
\be
\label{mehtaint}
\int \prod_{i=1}^N \rd\lambda_i |\Delta (\lambda)|^{2\beta} e^{-{1 \over 2} \sum_{i=1}^N \lambda^2_i} =(2\pi)^{N/2} \prod_{j=1}^N {\Gamma (1+ \beta j) \over \Gamma(1+ \beta)}. 
\ee
The result can be expressed in terms of the double Gamma Barnes function
\be
\Gamma_2(x|a,b)=\exp\left( {\rd \over \rd s}\Bigl|_{s=0} \zeta_2(s;a,b,x)\right), 
\ee
where the r.h.s. involves the Barnes double zeta function  
\be
\zeta_2(s;a,b,x)={1\over \Gamma(s)} \int_0^{\infty} \rd t \, t^{s-1} {\re^{-t x} \over (1-\re^{-a t}) (1-\re^{-b t})},
\ee
see \cite{spreafico} for a summary of properties of these functions. Indeed, it is easy to show that 
\be
 \prod_{j=1}^N \Gamma (1+ \beta j) =(2 \pi)^{N/2}  \beta^{N/2 +\beta N(N-1)/2} \Gamma(1+ N\beta) \Gamma(N) \Gamma_2^{-1} (N;1/\beta,1).
\ee
We can now obtain the large $N$ expansion of (\ref{betaz}) by using the asymptotic expansion of the Barnes double-Gamma function \cite{spreafico}, 
\be
\ba
\log \Gamma_2 (x;a,b)&={1\over ab} \left( -{1\over 2} x^2 \log x + {3\over 4} x^2\right) + {1\over2}\left( {a \over b} + {b \over a}\right) x \log x  -{1\over 12} \left( 2 + {a\over b} 
+{b \over a}\right) \log x \\
&-\chi'(0;a,b)+\sum_{n=3}^{\infty}  (n-3)! e_{n-2}(a,b) x^{2-n},
\ea
\ee
where $e_n(a,b)$ are defined by the expansion 
\be
{1\over (1-\re^{-at})(1-\re^{-bt})} =\sum_{n=-2}^{\infty} e_n(a,b)t^n
\ee
and $\chi(s;a,b)$ is the Riemann--Barnes double zeta function, 
\be
\chi(s;a,b)=\sum_{(m, n)\in \IN^2_0} (a m + b n )^{-s}, 
\ee
with $\IN^2_0=\IN^2\backslash \{ (0,0)\}$. 
Up to some additive terms, one finds 
\be
\ba
\label{fgas}
F&={1\over 2} \beta t^2 \left( \log (t) -{3\over 2}\right) g_s^{-2}, + {\beta-1\over 2} t  \left( \log(\beta t) -1\right)  g_s^{-1}  \\
&+{1-3 \beta +\beta^2\over 12 \beta} \log(\beta t) +{1-\beta \over 24  \beta t} g_s +{1-5\beta^2 +\beta^4 \over 720 \beta^3 t^2}  g_s^2+\cdots
\ea
\ee
where as usual 
\be
\label{thooft}
t=g_s N
\ee
is the 't Hooft coupling. From this expression we can read off the different $F_{k,l}$ of the Gaussian ensemble. The asymptotic expansion (\ref{fgas}) can be written as 
\be
F=- \log \Gamma_2\left( t; -g_s, {g_s/\beta} \right).
\ee
\end{example}

\subsection{The Chekhov--Eynard recursion for the beta ensemble}
\label{sec:ce}

When $\beta=1$, the full $1/N$ expansion (\ref{genusW}), (\ref{genusF}) of the matrix model was obtained in \cite{eynard,ceone} in terms of residue calculus on 
the spectral curve of the model. We recall that the spectral curve is defined by the following relation
\be
W_0(x)={1\over 2}\left( V'(x) -y(x)\right),
\ee
where $W_0(x)$ is the planar resolvent. 
In this paper we will be interested in the case of hyperelliptic spectral
curves. $y(x)$ can be written as 
\be
\label{scurve}
y(x)=M(x) {\sqrt{\sigma(x)}}, 
\ee
where 
\be
\sigma(x)=\prod_{i=1}^{2 s} (x-x_i), 
\ee
and thus realizes the plane complex curve $\Gamma = \{(x, y(x)), x\in \mathbb{C}\}$ as
a 2-sheeted cover of the complex plane, branched at $x=x_i$; if $p\in \Gamma$,
we will denote by $\bar p$ the conjugate point under the projection map to the
eigenvalue plane
\be (x(\bar p), y(\bar p)) = (x(p),-y(p)). 
\ee
In the following, we will often denote the eigenvalue location as $p$,
therefore writing $x(p)=p$ for the uniformization variable. The function $M(p)$ in (\ref{scurve}) is also called the {\it moment function}. 
In matrix models with polynomial potentials $M(p)$ is also a polynomial. If the potential contains simple 
logarithms, as in the Penner model that we will analyze later on, $M(p)$ is rather a rational function. In many situations related to 
topological string theory, $M(p)$ can be written in terms of an inverse hyperbolic function \cite{mmopen}. For future use, we will denote by $\CC$ a contour encircling the branch points 
and the branch cuts between them. \\

The Chekhov--Eynard recursion relation, proposed in \cite{ce}, gives a solution to the $1/N$ expansion (\ref{genusF}), (\ref{genusW}) in the general 
$\beta$ ensemble, in terms of period integrals 
defined on the spectral curve (\ref{scurve}). As
we will show in a moment, one important difference between the recursion proposed in
\cite{eynard,eo} and the one obtained in \cite{ce} is that, in the first case,
the recursion can be formulated in terms of residues in the branch points of
the curve.
However, in the recursion \cite{ce}, the expressions for $W_g$ with $g$ half-integer involve contour integrals
where the integrand has branch cuts, and they can {\it not} be reduced to residues
at the branch points.

The starting point to derive the recursion relations are the loop equations of the $\beta$ ensemble. In the following we will assume that $V(p)$ is a polynomial 
of degree $d$. The loop equations have been written down explicitly in \cite{em}, and they 
read, with the notations above, 
\be
\ba
&V'(p_1) W (p_1, \cdots, p_k)-U(p_1, \cdots, p_k)\\
&=2 W(p_1) W(p_1, \cdots, p_k) + \hbar^2 W(p_1, p_1, \cdots, p_k) + \hbar \gamma {\partial \over \partial p_1}W (p_1, p_2, \cdots, p_k) \\
&+ \sum_{j=1}^{k-2} \sum_{I\in K_j} W(p_1, p_I) W(p_1, p_{K\backslash I}) \\
&+\sum_{j=2}^k {\partial \over \partial p_j} {W(p_2, \cdots, p_j, \cdots, p_k) - W(p_2, \cdots, p_1, \cdots, p_k) \over p_j -p_1},
\ea
\ee
for $k\ge 2$, while for $k=1$ we have simply
\be
V'(p) W (p)-U(p)= W^2(p) + \hbar^2 W(p, p) + \hbar \gamma {\partial \over \partial p}W (p).
\ee
In these equations, $U(p_1, \cdots, p_k)$ is a polynomial in $p_1$ of degree $\delta_{k2}+ {\rm deg}(V')-2$. 
It turns out that these equations can be solved recursively in the $g_s$ expansion. 
To see this, let us look at the simple example of $k=1$, and let us plug in
the expansion (\ref{genusW}). The first $\beta$-ensemble correction is 
$W_{1/2}(p)$. It satisfies the equation
\be
\label{loopeq}
\left(V'(p)-2 W_0(p)\right) W_{1/2}(p)-U_{1/2}(p)=\gamma {\partial W_0(p) \over \partial p}.
\ee
This can be solved as,  
\be
{\sqrt{\sigma(p)}} W_{1/2}(p)={\gamma \over M(p)}  {\partial W_0(p) \over \partial p} + {U_{1/2}(p) \over M(p)}.
\ee
Notice that the r.h.s. in this equation is {\it not} a rational function, as it happens in the solution of the loop equations in the $\beta=1$ case, since the 
derivative of the planar resolvent involves the multivalued function ${\sqrt{\sigma(p)}}$. However, one can still use 
the techniques developed in \cite{eynard,ceone,eoreview} in order to give an explicit expression for $W_{1/2}(p)$. 
Let $\rd S (p,q)$ denote the unique third kind differential on the 
spectral curve having a simple pole at $p=q$ and $p=\bar q$
with residues $+1$ and $-1$ respectively and vanishing $A$-periods. We can write 
\be
W_{1/2}(p_1) =-{\rm Res}_{p=p_1} \rd S (p_1,p) W_{1/2}(p),
\ee
where $p_1$ is a point outside $\CC$. We now take into account that $W_{1/2}(p)$ has no residues at points away from the contour 
$\CC$, as well as no residue at $p=\infty$. The first 
fact follows from the assumption that there are no eigenvalues of the matrix model away from the cut (see \cite{eynard}, eq. (2.13)), and the second 
fact follows from the expansion at infinity expressing $W_{1/2}(p)$ in terms of correlation functions, 
\be
W(p) =g_s \sum_{n\ge 1}{\langle \tr \, M^n\rangle \over p^{n+1}}.
\ee
By contour deformation, we find that 
\be
W_{1/2}(p) ={1\over 2 \pi \ri} \oint_{\CC} \rd S (p,q) W_{1/2}(q).
\ee
Using now the loop equation and the expression for the spectral curve, we find 
\be
W_{1/2} (p) ={1\over 2 \pi \ri}\oint_{\CC} {\rd S (p, q) \over y(q)} \gamma {\partial \over \partial q} W_0(q)
+\oint_{\CC} {\rd S (p, q) \over y(q)} U_{1/2}(q).
\ee
Since $U_{1/2}(q)$ is a polynomial in $q$, the last integral vanishes, and we obtain
\be
\label{onepcorrection}
W_{1/2} (p) ={1\over 2 \pi \ri} \oint_{\CC} {\rd S (p, q) \over y(q)} \left[ \gamma {\partial \over \partial q} W_0(q) \right].
\ee
The same result can be obtained using the inversion operator of
\cite{ceone,ce}. Using the expansion (\ref{ggamma}), we can rewrite this as 
\be
W_{0,1}(p)=-{1\over 4 \pi \ri} \oint_{\CC} \rd S (p, q) { y'(q) \over y(q)}, 
\label{01for}
\ee
where we have assumed that $V''(q)$ is analytic inside $\CC$. 

Let us now consider the case $k=2$. The first non-trivial correction in the $\beta$ ensemble to the two-point function is $W_{1/2}(p_1, p_2)$. It satisfies the equation
\be
\ba
& \left( V'(p_1) -2 W_0 (p_1)\right) W_{1/2} (p_1, p_2) -U_{1/2}(p_1, p_2) \\
&=2 W_{1/2}(p_1) W_0(p_1, p_2) + \gamma {\partial \over \partial p_1} W_0(p_1, p_2) + {\partial \over \partial p_2} \left[ {W_{1/2}(p_2) -W_{1/2}(p_1)\over p_2-p_1}\right]. 
\ea
\ee
We can use the same contour deformation argument. There will not be any contribution from the polynomial $U_{1/2}(p_1, p_2)$ nor from 
\be
{\partial \over \partial p_2} \left[ {W_{1/2}(p_2)\over p_2-p_1}\right].
\ee
However, there is a contribution from 
\be
-{\partial \over \partial p_2} \left[  {W_{1/2}(p_1)\over p_2-p_1}\right]={W_{1/2}(p_1)\over (p_1-p_2)^2}
\ee
and the final expression is, 
\be
W_{1/2} (p_1, p_2) =\oint_{\CC} {\rd S (p_1, p) \over y(p)} \biggl[ 2 W_{1/2}(p_1) \left( W_0(p_1, p_2) +
{1\over 2} {1\over (p_1-p_2)^2} \right) + \gamma {\partial \over \partial p_1} W_0(p_1, p_2) \biggr].
\ee
It involves the ``corrected" two-point function as in \cite{eynard} and subsequent works\footnote{This correction does not appear in the formulae of \cite{ce}, see \cite{chekhov} for a careful statement of the recursion.}. Notice again that the integrand in the above formula is not a rational 
function, due to the derivative term. 

One can see that the general solution for the ``genus" $g$ correlators $W_g(p_1, \cdots, p_h)$ is 
\be
 \label{recursive}
\ba
 \CW_g (p,p_1,\dots,p_k) 
&=\oint_{\CC} {\rd q\over 2\pi \ri} {\rd S(p,q)\over y(q)}\,\biggl(\sum_{h=0}^g \sum_{J\subset K} \CW_h(q,p_J)\CW_{g-h}(q,p_{K/J}) \\
& \qquad + \CW_{g-1}(q,q,p_K) + \gamma {\rd \over \rd q} \CW_{g-1/2}(q, p_1, \cdots, p_k) \biggr)
\ea
\ee
where 
\be
\CW_g(p_1,\dots,p_k) =W_g(p_1,\dots,p_k) +{1\over 2} {\delta_{k2}\delta_{g0} \over  (p_1-p_2)^2}.
\ee

The free energies can be computed by using the loop inversion operator introduced in \cite{ceone,ce,eo}. In the ``stable" case, i.e. for $(k,l)\not=(0,0)$, $(0,1)$, $(1,0)$ and $(0,2)$, they are 
given by 
\be
F_{k,l}={1\over 2-2k-l}\oint_{\CC} {\rd q\over 2\pi \ri} \Phi(q) W_{k,l}(q), 
\ee
where
\be
\Phi'(q)=y(q)
\ee
is a primitive of the spectral curve. For the unstable cases, we have specific formulae which can be found in \cite{ce,chekhov}. In this paper we will be particularly interested in the first correction to 
the free energy, which is given by 
\be
\label{f01}
F_{0,1}= {1\over 2\pi} \int_{\CC} \rd q |y(q)| \log |y(q)|. 
\ee
Here, the integration is over the union of the intervals where the density of eigenvalues is non-vanishing. To make our notation simpler, we have denoted 
this support by $\CC$ again. 

In order to obtain concrete results for the correlators using (\ref{recursive}), we need explicit formulae for the differential $\rd S(p,q)$. 
When the spectral curve is of the form (\ref{scurve}) we can proceed as follows \cite{eynard}. We define the $A_j$ cycle of this curve 
as the cycle around the cut
\be
(x_{2j-1}, x_{2j}), \qquad j=1, \cdots, s-1.
\ee
There exists a unique set of $s-1$ polynomials of degree $s-2$, denoted by $L_j(p)$,
such that the differentials 
\be
\omega_j ={1\over 2\pi \ri} {L_j(p)\over \ssq{p}}\rd p
\ee
satisfy
\be
\oint_{A_j}  \omega_i = \delta_{ij}, \qquad i,j=1, \cdots, s-1.
\ee
The $\omega_i$s are called normalized holomorphic differentials. The differential $\rd S(p,q)$ can then be written as
\be
\label{CjLambda}
\rd S (p,q)=
{\ssq{q}\over \ssq{p}}\,\left(
{1\over p-q}-\sum_{j=1}^{s-1} C_j(q)L_j(p)
\right)\, \rd p
\ee
where
\be\label{defCj}
C_j(q):={1\over 2\pi \ri }\,\oint_{A_j} {\rd {p}\over {\sqrt{\sigma(p)}}}\,{1\over p-q}.
\ee
In this formula, it is assumed that $q$ lies outside the contours $A_j$.
One has to be careful when $q$ approaches some branch point $x_j$.
When $q$ lies inside the contour $A_j$, then one has:
\be
\label{regc}
C^{\rm reg}_l(q)+{\delta_{lj}\over \ssq{q}}={1\over 2\pi \ri}\,\oint_{A_j} {\rd {p}\over \ssq{p}}{1\over p-q}
\ee
which is analytic in $q$ when $q$ approaches $x_{2j-1}$ or $x_{2j}$. 
%
%
%

\subsection{One-cut examples} 
In the one-cut case we simply have
\be
{\rd S (p,q)\over y(q)}={1\over M(q) {\sqrt{\sigma(p)}} (p-q)}.
 \ee
 We will now present some explicit formulae for the very first corrections to the connected correlators. 
\\ 

  The first correction to the resolvent is given by (\ref{w01}), and we find 
 \be
 \label{w01}
W_{0,1}(p)=-{1\over 2 {\sqrt{\sigma(p)}}}\oint_{\CC} {\rd q \over 2 \pi \ri} {y'(q) \over M(q)  (p-q)}.
\ee
An explicit, general formula for this correlator was obtained in \cite{gm} by using contour deformation. Assuming we have a polynomial potential of degree $d$, 
we will write the moment function as
\be
\label{momentf}
M(z)=c \prod_{i=1}^{d-2} (z-z_i)
\ee
where $c$ is a constant. We can calculate (\ref{onepcorrection}) by deforming the contour. This picks a pole at $q=p$, a pole at infinity, and poles at the zeroes 
of $M(z)$. A simple computation gives 
\be
\label{simpleonecut}
 W_{0,1}(p)= -{1\over 2} {y'(p) \over y(p)} +{1\over 2 {\sqrt{\sigma(p)}}} \biggl[ d-1 +\sum_i {\sqrt{\sigma(z_i)} \over p-z_i}\biggr].
 \ee
 This can be written in a way which makes manifest the absence of singularities at $p=z_i$:
\be
\label{exonecc}
W_{0,1}(p)={d-1\over 2{\sqrt{\sigma(p)}}} -{1\over 4} {2p-a-b\over (p-a)(p-b)}-{1\over 2 {\sqrt{\sigma(p)}}}\sum_i \biggl[ {{\sqrt{\sigma(p)}} -{\sqrt{\sigma(z_i)}} \over p-z_i} \biggr].
\ee

In the one-cut case it is also possible to write a very explicit formula for $F_{0,1}$ (or rather for its derivative w.r.t. the 't Hooft parameter $t$). 
Using that (see for example \cite{dfgzj})
\be
\label{deryt}
\partial_t y(q)=-{2\over {\sqrt{\sigma(q)}}}
\ee
we find 
\be
\label{derexf}
\partial_t F_{0,1}=1 +{1\over  \pi}  \int_{\CC} \rd q { \log |y(q)| \over {\sqrt{|\sigma(q)|}}}.
\ee
This is easy to calculate in terms of the parameters (\ref{momentf}) appearing in the moment function, and one finds the general one-cut expression, 
\be
\partial_t F_{0,1}=1 + {1\over 2} \log \Bigl( {b-a\over 4}\Bigr)^2+ \log c + \sum_i \log \biggl[ {1\over 2} \Bigl(z_i -{a+b\over 2}  + {\sqrt{\sigma(z_i)}}\Bigr) \biggr].
\ee
For higher corrections, general formulae become cumbersome (see \cite{eynardformulae} for an example), but expressions for particular potentials are easy to 
derive. 

\begin{example} {\it The Gaussian potential}. Let us consider the Gaussian potential,
 \be
V(x)= {x^2\over 2}.
 \ee
In this case, the moment function $M(p)$ is trivial and we simply obtain
\be
W_{0,1}(p)={W_0'(p)\over y(p)} ={1\over 2}\biggl({1 \over {\sqrt{p^2-4t}}}-{p\over p^2-4t}\biggr). 
\ee
Higher order correlators can be similarly computed in a straightforward
fashion from \eqref{recursive}. We find for example\footnote{Previous arXiv versions
of this paper contained an erroneous expression for $W_{1,2}(p)$ in \eqref{eq:w12}; this came to
our attention after the appearance of \cite{Witte:2013cea}.}
\ben
W_{0,2}(p) &=& \frac{p^2+t}{\left(p^2-4 t\right)^{5/2}}-\frac{p}{\left(p^2-4
  t\right)^2},  \\
W_{0,3}(p) &=& 5\l(\frac{ p^2+t}{\left(p^2-4 t\right)^{7/2}}-\frac{ p^3+2 p t}{\left(p^2-4 t\right)^4}\r), \\
W_{1,1}(p) &=& \frac{1}{2} \left(\frac{p^2+6 t}{\left(p^2-4
  t\right)^{7/2}}-\frac{p \left(p^2+30 t\right)}{\left(p^2-4
  t\right)^4}\right),
  \\
W_{1,2}(p) &=& \frac{1}{2}\l(\frac{23 p^4+454 p^2 t+176 t^2}{ \left(p^2-4
  t\right)^{11/2}}-\frac{ 23 p^3+180 p t}{ \left(p^2-4
  t\right)^5}\r).
\label{eq:w12}
\een

\end{example}

\begin{example} {\it The cubic potential}. Let us consider a cubic potential 
\be
V(x)={x^2\over 2}+{g\over 3} x^3,
\ee
with a classical maximum at
$p=-1/g$ and a minimum at $p=0$ (see \figref{fig:cubonecut}).
\begin{figure}
\centering
\includegraphics[scale=0.35]{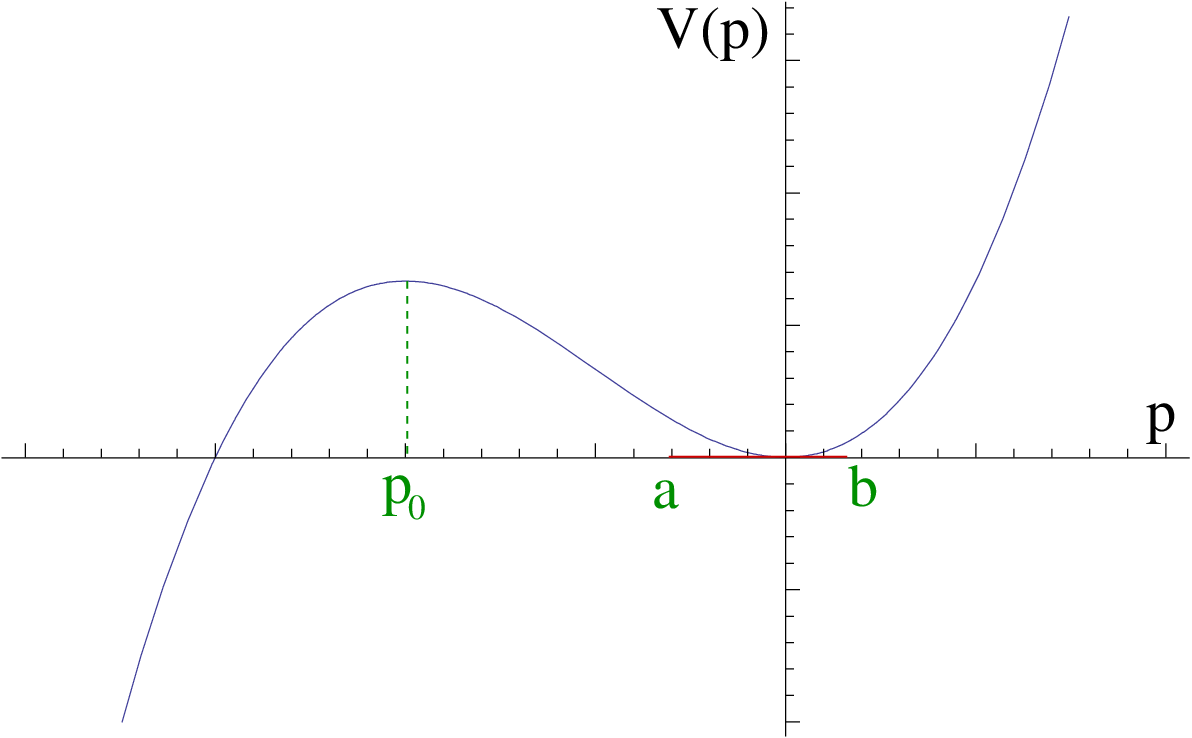}
\caption{The cubic matrix model in the single-cut phase.}
\label{fig:cubonecut}
\end{figure}
In the stable
one-cut phase the eigenvalue density is supported on an interval $(a, b)$ around $p=0$:
the spectral curve then takes the form
\be
y(p)=M(p) \sqrt{(p-a)(p-b)}, \qquad M(p)=g(p-p_0) 
\ee
where
\be
p_0=-{1\over g} -{a+b\over 2}.
\ee
The branch points can be expressed as a
function of the (only) 't Hooft parameter by imposing the correct asymptotics
for the planar resolvent \cite{bipz}, and one finds, as a power series in $t$, 
\ben
a &=& 2 \sqrt{t}-2 g t+4 g^2 t^{3/2}-12 g^3 t^2+36 g^4 t^{5/2}-128 g^5
t^3+\CO\left(t^{7/2}\right), \nn \\
b &=& -2 \sqrt{t}-2 g t-4 g^2 t^{3/2}-12 g^3 t^2-36 g^4 t^{5/2}-128 g^5
t^3+\CO\left(t^{7/2}\right).\nn
\een
It is now straightforward to compute $\beta$-deformed correlators from
\eqref{recursive}. For example, \eqref{exonecc} 
gives
%
\ben
W_{0,1}(p) &=& -\frac{a (2 b-3 p+p_0)-3 b p+b p_0+4 p^2-2 p
  p_0}{4 \sigma(p) (p-p_0)}-\frac{\sqrt{(a-p_0)
    (b-p_0)}-2 p+2 p_0}{2 (p-p_0) \sqrt{\sigma(p)}}. \nn \\
\een
As an instance we have, up to order $t^3$ and $g^7$
\ben
 \bra \tr M^3 \ket &=& \frac{-4 g t^3 + \ldots}{g_s}+ \l(1-\beta^{-1} \r) \left(9 g t^2+118 g^3
  t^3 + \dots \right)\nn \\  &+&  g_s \beta^{-1} \left (g t+28 g^3 t^2 + 664 g^5
t^3 + \dots\right) \nn \\ &+&  g_s(1-\beta^{-1})^2 \l(\frac{6}{g}+17 g t + 182 g^3 t^2+2228 g^5
   t^3 + \dots\r) \nn \\ &+& g_s^2(1-\beta^{-1}) \left(3 g+198 g^3 t+6959 g^5 t^2+202254 g^7 t^3+\dots\right)
+\dots    \\  
\bra \tr M^2 \tr M \ket^{(c)} &=&
-4 \l(g t^2+10 g^3 t^3+\ldots \r)+ g_s \l(1-\beta^{-1}\r)\left(\frac{2}{g}+8 g t+106 g^3 t^2+1640
   g^5 t^3 \right)+\dots \nn \\
\een
\end{example} 

\begin{example} {\it The quartic potential}. Consider finally a potential of the form
\be
\label{quartpot}
V(x)={x^2 \over 2} +g x^4.
\ee
 The resolvent is given by \cite{bipz}
\be
\label{quarticres}
W_0(z)={1\over 2}\Bigl( z + 4 g z^3 -\bigl( 1+ 8 ga^2 + 4 g z^2\bigr){\sqrt {z^2-4 a^2}}\Bigr),
\ee
where $a$ is a function of $g,t$
\be
\label{asq}
a^2={1\over 24 g} \biggl( -1+ {\sqrt { 1 + 48 g t}}\biggr).
\ee
 The moment function has two zeros at 
\be
z_0^2=-{1+ 8 g a^2\over 4g}, 
\ee
and (\ref{exonecc}) gives \cite{gm}
\be
W_{0,1}(z)=-{1\over 2} {z\over z^2-4 a^2} +{3 \over 2 z\sqrt{ 1-4a^2/z^2}}+ {\sqrt{ 1-4a^2/z_0^2 \over 1-4 a^2/z^2}} {z_0^2 \over z(z^2-z_0^2)}-{z\over z^2-z_0^2}.
   \ee
This expression leads to explicit results for the enumeration of quadrangulations of the projective plane $\IR \IP^2$, see \cite{gm} for more details. 
\end{example}

\subsection{Two-cut examples}
\label{sec:twocut}
Let us now consider the two-cut case, where we have $s=2$. In this elliptic case there is one single integral $C_1(p)$ \eqref{defCj} to 
compute, and we can obtain very explicit expressions in terms of elliptic integrals \cite{bkmp}:
\be
\label{genintegral}
\ba
C_1(p)&= {2\over \pi (p-x_3) (p-x_2) {\sqrt{(x_1-x_3)(x_2-x_4)}}} 
\biggl[ (x_2-x_3) \Pi(n_4, k) + (p-x_2) K(k)\biggr], \\
C^{\rm reg}_1(p)&=  {2\over \pi (p-x_3) (p-x_2) {\sqrt{(x_1-x_3)(x_2-x_4)}}} 
\biggl[ (x_3-x_2) \Pi(n_1, k) + (p-x_3) K(k)\biggr], \\
L_1 &= \frac{\pi  \sqrt{(x_1-x_3) (x_2-x_4)}}{2 K(k)},
\ea
\ee
where 
\be
k^2 ={(x_1-x_2)(x_3-x_4)\over(x_1-x_3)(x_2-x_4)}, \qquad n_4= {(x_2-x_1)(p-x_3) \over (x_3-x_1)(p-x_2)}, \qquad n_1= {(x_4-x_3)(p-x_2) \over (x_4-x_2)(p-x_3)},
\label{eq:kn4n1}
\ee
$\Pi(n, k)$ is the elliptic integral of the third kind,
\be
\Pi(n, k)=\int_0^1 {\rd t\over (1-n t^2) {\sqrt {(1-t^2)(1-k^2 t^2)}}}
\ee
and $K(k)$ is the standard elliptic integral of the second kind. The leading correction $W_{0,1}$
to the resolvent is given by (\ref{w01}). We will split the r.h.s as $ W^{(A)}_{0,1}(p)+ W^{(B)}_{0,1}(p)$, where
\ben
\label{eq:omegaA}
W^{(A)}_{0,1}(p) & := & -\frac{1}{4\pi {\rm i} \ssq{p}}\oint_\CC \frac{y'(q) \rd q}{M(q)(p-q)},
\\
W^{(B)}_{0,1}(p) & := & \frac{1}{4\pi {\rm i}\ssq{p}}\l[ \oint_{\CC_1}
  \frac{C^{\rm reg}_1(q) L_1 y'(q) \rd q}{M(q)}+\oint_{\CC_2} \frac{C_1(q) L_1
    y'(q)
    \rd q}{M(q)}\r].
\label{eq:omegaB}
\een
When $M(p)$ is a rational function of $p$, the integrand in \eqref{eq:omegaA} is a single valued meromorphic function
outside the cuts and we can compute $W^{(A)}_{0,1}(p)$ by deforming the
contour and picking up poles just as we did for the single cut case. On the
other hand, as was pointed out in the discussion of Section \ref{sec:ce},
this is not the case for the expressions \eqref{genintegral} for $C_1(p)$ and
$C_1^{\rm reg}(p)$, which are only well-defined in the neighbourhood of the cuts $[x_3, x_4]$ and $[x_1, x_2]$
respectively. A way to treat the integrals appearing in \eqref{eq:omegaB} is
the following: for a fixed polarization of the spectral curve, the elliptic
modulus  $k$ in \eqref{eq:kn4n1} vanishes by definition when we shrink the
$A$-cycle. By expanding the complete elliptic integrals $\Pi(n,k)$ and
$K(k)$ appearing in \eqref{genintegral} around $k=0$ and integrating term by
term, we obtain an expansion of the form
\be
W^{(B)}_{0,1}(p) =  \frac{1}{4\pi {\rm i}\ssq{p}} \sum_{n=0}^\infty \l[
  \oint_{\CC_1} \frac{C^{[n], \rm reg}_1(q) L_1 y'(q) \rd q}{M(q)}+\oint_{\CC_2}
  \frac{C^{[n]}_1(q) L_1y'(q) \rd q}{M(q)}\r] k^n,
\label{eq:kexpomegaB}
\ee
where we denoted 
\be
f^{[n]} := \frac{1}{n!}\frac{\de^n f}{\de k^n}\Bigg|_{k=0}.
\ee
At any fixed order in $k$, by formulae \eqref{eq:expk} and \eqref{eq:exppi}, the
integrands of \eqref{eq:kexpomegaB} are algebraic functions of $q$ as long as
the moment function is rational, and can be
computed exactly in terms of complete elliptic integrals. 

It should be stressed that, while \eqref{eq:kexpomegaB} yields only a perturbative
expression valid for small $k$, this procedure holds true for a {\it generic},
fixed choice of polarization\footnote{In particular, it continues to hold true when we
vary the choice of $A$ and $B$ cycles, thereby
changing the very definition of $\rd S(p,q)$ and $k$; for example, in the context of
Seiberg-Witten curves, this would allow us to find expansions in any $S$-duality
frame, also at strong coupling.}. It therefore provides a way to
expand the amplitudes around any boundary point in the moduli space where the spectral
curve develops a nodal singularity. 
 
\begin{example} {\it{The cubic matrix model}}. As a first application of our formulae, let us consider the case of the cubic
matrix model with 
\be
V(p)={p^2 \over 2} +  g {p^3  \over 3}
\ee
in the two-cut case. The spectral curve reads
\be
y(p)=M(p) \ssq{p}, \quad M(p)=g, \quad \sigma(p)=\sqrt{(p-x_1)(p-x_2)(p-x_3)(p-x_4)}.
\ee
Following \cite{civ, kmr} we can parametrize the branch points in terms of a
pair of ``B-model'' variables ($z_1$, $z_2$) as
\be
\sum_i x_i =2 Q, \quad x_2-x_1=2 \sqrt{z_1}, \quad x_4-x_3=2 \sqrt{z_2}, \quad -x_1-x_2+x_3+x_4=2 I
\ee
where
\be
Q=-\frac{1}{g}, \quad I=\sqrt{\frac{1}{g^2}-2 (z_1+z_2)}.
\ee
The 't Hooft parameters can be computed explicitly in terms of complete
elliptic integrals \cite{khuang} as
\ben
t_1 &=& \frac{(x_4-x_3)(x_2-x_4)(x_1-x_4)^2}{\pi\sqrt{(x_1-x_3)(x_2-x_4)}}\Pi\l(\tilde n_1,k\r), \\
t_2 &=& \frac{(x_4-x_2)(x_2-x_1)(x_3-x_2)^2}{\pi\sqrt{(x_1-x_3)(x_2-x_4)}}\Pi\l(\tilde n_2,k\r),
\label{eq:ff2cutcub}
\een
where
\be
k^2=\frac{(x_1-x_2) (x_3-x_4)}{(x_1-x_3) (x_2-x_4)}, \quad \tilde n_1 = \frac{x_3-x_4}{x_3-x_1}, \quad \tilde n_2=\frac{x_1-x_2}{x_1-x_3}.
\ee
This can be inverted as
\ben
z_1 &=& -4
t_{1} + 16 g^2 t_{1}^2-24 g^2 t_{1} t_{2} + \dots,\\
z_2 &=& 4 t_{2} 
-24 g^2 t_{1} t_{2}+16 g^2 t_{2}^2 + \dots
\een
Let us turn to compute $W^{(B)}_{0,1}(p)$ first. We can write it as
\be
W^{(B)}_{0,1}(p)=\frac{I_1(t_1, t_2) + I_2(t_1,t_2)}{\ssq{p}}
\label{eq:omegaBcub}
\ee
with $$I_j = \sum_{n=0}^\infty \l[\oint_{A_j} \frac{C^{[n], \rm reg}_1(q)
    y'(q)}{4 \pi i M(q) L}\r] k^n.$$ We find
\ben
-I_1 &=& \frac{1}{32} z_{1} \left(8 g+24 g^3 z_{2}+81 g^5 z_{2}^2\right)+\frac{1}{64} z_{1}^2 \left(20
   g^3+208 g^5 z_{2}+1269 g^7
   z_{2}^2\right)+ \dots \nn
   \\
I_2 &=& \left(\frac{g z_{2}}{4}+\frac{5 g^3 z_{2}^2}{16}\right)+z_{1} \left(\frac{3 g^3
   z_{2}}{4}+\frac{13 g^5 z_{2}^2}{4}\right)+z_{1}^2 \left(\frac{81 g^5
  z_{2}}{32}+\frac{1269 g^7 z_{2}^2}{64}\right)+\dots \nn \\
\een
whereas the residue computation for $W^{(A)}_{0,1}(p)$ yields
\be
W^{(A)}_{0,1}(p) = \frac{1}{4}
\sum_{i=1}^4 \frac{1}{x_i-p}
+\frac{1+2
g p}{2 g \ssq{p}}.
\label{eq:omegaAcub}
\ee
We can compare our result to explicit perturbative computations for the
$\beta$-deformed cubic matrix model, along the lines of \cite{kmt}. As an
example, \eqref{eq:omegaBcub}, \eqref{eq:omegaAcub} together yield up to
quadratic order in $t_1$ and $t_2$
\ben
\bra \tr M \ket &=& \left(  -\frac{t_{1}}{g}-g t_{1}^2-gt_{2}^2+4 g t_{2}t_{1} +30 g^3
   t_{2}^2t_{1}  -30 g^3 t_{2}t_{1}^2-708 g^5
   t_{2}^2t_{1}^2+ \dots\right) g_s^{-1} 
\nn \\ &+& \l(1-\beta^{-1}\r)\l[
g (t_{1}+t_{2}) + g^3 \left(9 t_{2}^2-9 t_{1}^2\right)+g^5 \left(-162 t_{1}^2 t_{2}-162 t_{1}
t_{2}^2\r) + \dots \r] + \dots, \nn \\
\een
which perfectly agrees with the computation from perturbation theory. \\

Interestingly, a closed form expression for $W_{0,1}(p)$ can be
found as a function of the branch points. It was shown in \cite{shigemori2}
that, for matrix models with constant moment function $M(p)=g$, $W_{0,1}$
is directly related to the planar resolvent as follows
\be
W_{0,1}(p)=\de_t W_{0}(p) - \frac{1}{4}\de_p \ln \sigma(p).
\label{eq:shigemori}
\ee
The first term of the r.h.s. can be evaluated very explicitly upon
expressing the derivative w.r.t. the total 't Hooft coupling in terms of
derivatives with respect to the branch points, following \cite{msw}. The partial derivatives $A_{i,j}=\frac{\de x_i}{\de t_j}$ satisfy the linear system
\ben
\label{eq:syscub1}
\sum_{i=1}^4 M(x_i) x_i^k A_{i,j} &=& 4 \delta_{k,2},  \\
\sum_{i=1}^4 M(x_i) K_i A_{i,j} &=& 4 \pi \delta_{j,2},
\label{eq:syscub2}
\een
where we denoted
\be
K_i=\int_{x_3}^{x_4} \frac{\ssq{p}}{p-x_i}.
\ee
As the $A_{i,j}$ are completely determined by \eqref{eq:syscub1}-\eqref{eq:syscub2}, it is straightforward to perform explicitly 
the derivatives in \eqref{eq:shigemori} and obtain a compact expression for $W_{0,1}(p)$ as a function of the branch points. We get
\be
W_{0,1}(p) =
\frac{1}{\ssq{p}}\Bigg[(x_2-x_3)  \frac{\Pi \left(\tilde n_2, k\right)}{K\left(k\right)}  
- \frac{\pi  \sqrt{(x_1-x_3) (x_2-x_4)}}{4 K\left(k\right)} +p-x_3 \Bigg] -\frac{\sigma'(p)}{4\sigma(p)}.
\label{eq:w01cub}
\ee
It is worthwhile to remark that this expression has a more involved dependence
on the branch points as compared to oriented, open string amplitudes at higher
genus. The ordinary topological recursion \cite{eynard, eo} prescribes the following
general form for the $\beta=1$ correlators in the two-cut case
\be
W_{g,0}(p_1, \dots, p_h)=\sum_{n=0}^{3g-3+2h} \l(\frac{E(k)}{K(k)}\r)^n f_n(\{x_i\}, \{p_j\})=\sum_{n=0}^{3g-3+2h} \l(E_2(\tau)\r)^n \tilde{f}_n(\tau, \{p_j\})
\ee
where $\tau$ is the half-period ratio on the mirror curve, $\tilde f_n$ are
holomorphic, weight $-2n$ modular forms for fixed $p_j$, and $E_2(\tau)$ is
the second Eisenstein series (see \cite{bkmp2, bt} for a detailed
discussion). In particular, only first- and second-kind elliptic integrals are
involved for $\beta=1$, whereas in the $\beta$-deformed case, as \eqref{eq:w01cub}
shows, we have a more sophisticated dependence on closed string moduli due to
the appearance of elliptic integrals of the third kind at prescribed values
for the elliptic characteristic.  It would be interesting to track the origin
of this higher degree of complexity for $\beta$-deformed amplitudes.

\end{example}

\begin{example} {\it The symmetric double-well}. As the simplest instance of a two-cut model with 
non-trivial moment function,
consider the double well potential 
\be
V(p)=-{p^2\over 2} + g {p^4 \over 4},
\ee
depicted in
Fig. \ref{fig:doublewell}. 
\begin{figure}
\centering
\includegraphics[scale=0.35]{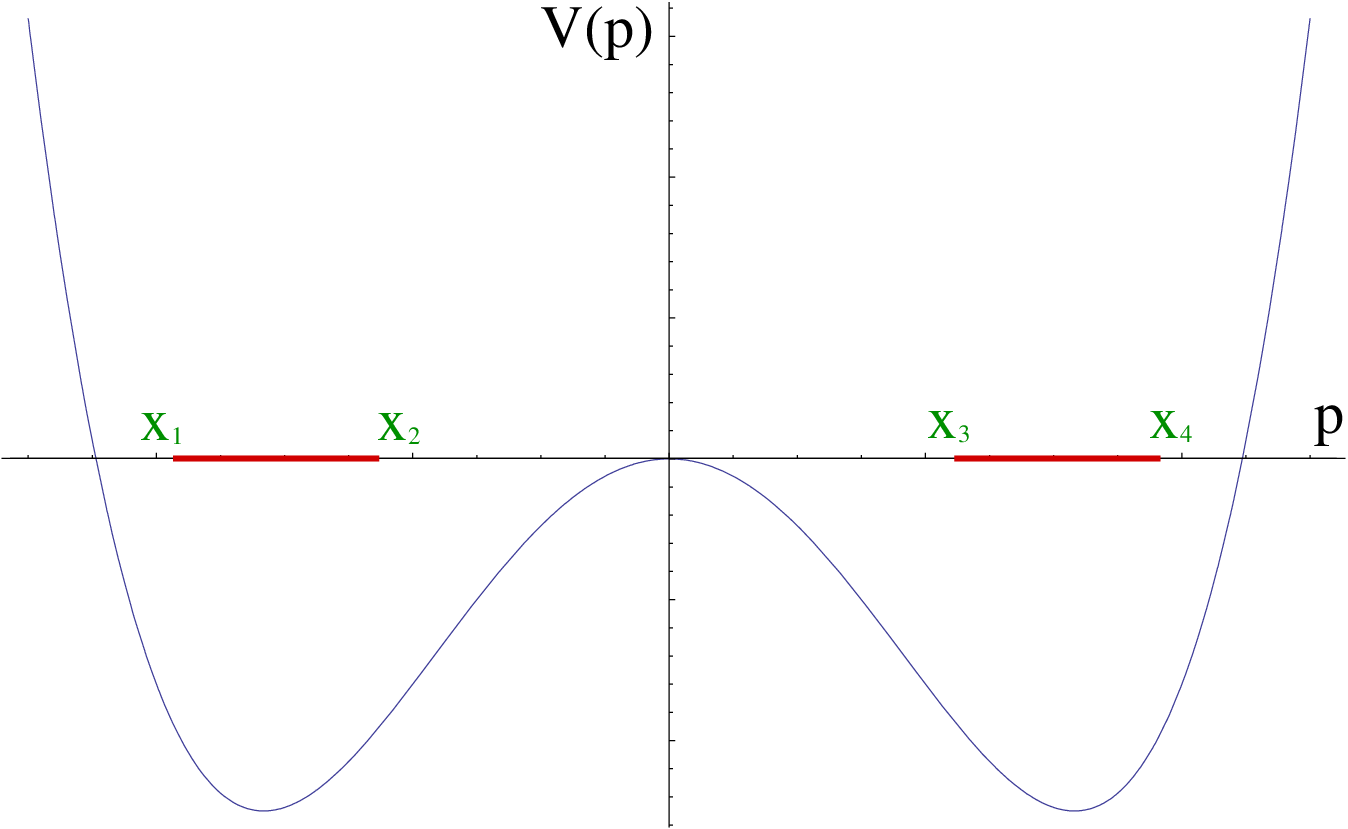}
\caption{The double-well potential in the symmetric 2-cut phase.}
\label{fig:doublewell}
\end{figure}
The potential has two minima at $p=\pm 1/\sqrt{g}$ and a
maximum at $p=0$. For simplicity we consider the case in which we
equally distribute the eigenvalues between the two minima, i.e. we restrict to the symmetric slice
$t_1=t_3=t/2$, $t_2=0$. The moment function in this case takes the form
\be
M(p)= g p.
\ee
The branch points can be readily computed as a function of the total 't Hooft
coupling $t$ by imposing the $\mathbb{Z}_2$ symmetry between the cuts and the
leading asymptotics of the resolvent. We get
\be
x_1 = -\sqrt{\frac{1}{g}+\frac{2 \sqrt{t}}{\sqrt{g}}}, \quad x_2 =
-\sqrt{\frac{1}{g}-\frac{2 \sqrt{t}}{\sqrt{g}}}, \quad x_3 =
\sqrt{\frac{1}{g}-\frac{2 \sqrt{t}}{\sqrt{g}}}, \quad x_4 = \sqrt{\frac{1}{g}+\frac{2 \sqrt{t}}{\sqrt{g}}}.
\label{eq:bpts2well}
\ee
We now turn to compute $W^{(B)}_{0,1}(p)$. In this case, the integrals in \eqref{eq:omegaB} can be computed exactly. To see this, let us consider the $PSL(2, \mathbb{C})$ transformation
\be
p \to \tilde p =\frac{\alpha p+ \beta}{\gamma p + \delta}, \quad A=\left(
\begin{array}{ll}
\alpha & \beta \\ \gamma & \delta
\end{array}
\right)
\in PSL(2,\mathbb{C})
\ee
with
\be
\ba
\alpha&=-\delta= \l(x_4x_1-x_2x_3\r) \zeta, \\
 \beta &= (x_1 x_2 x_3-x_1 x_3 x_4+x_2 x_3 x_4-x_1 x_2 x_4)\zeta, \\
\gamma&=(x_1+x_4-x_2-x_3)\zeta.
\ea
\ee
In these equations, $\zeta$ is given by
\be
\zeta := \frac{(x_4-x_1) }{(x_1-x_3)\sqrt{(x_1-x_4)
    (x_1-x_2)(x_2-x_4)(x_3-x_4)}}.
\ee
Let us apply this transformation to the second integral on the r.h.s. of \eqref{eq:omegaB}. Then the sum of the two integrals becomes the single definite integral
\be
\oint_{\CC_1} \frac{\rd q}{4 \pi \ri}\l[\l(\frac{y'(\tilde q)}{M(\tilde
    q)}+\frac{y'(q)}{M(q)}\r) C_1^{\rm reg}(q) L_1 + \frac{ (x_1-x_2-x_3+x_4) y'(\tilde q)}{ (x_1 (x_4-q)+(x_3-x_4) q+x_2 (q-x_3)) M(\tilde q) }  \r].
\ee
In the case of the symmetric double-well, \eqref{eq:bpts2well} implies
\be
\frac{y'(\tilde q)}{M(\tilde q)}=-\frac{y'(q)}{M(q)}, \quad x_1+x_4-x_2-x_3=0,
\ee
therefore
\be
W^{(B)}_{0,1}(p)=0.
\ee
%
For
$W^{(A)}_{0,1}(p)$ we instead find
\be
W^{(A)}_{0,1}(p)=\frac{\sqrt{1-4 t}+3 g p^2}{2 p g \ssq{p}}+\frac{-3 g^2 p^4+g
  \left(4 p^2+4 t\right)-1}{2 p g^2 \sigma(p)}.
\ee
As an example, this yields
\be
\bra \tr M^4 \ket_{0,1}=
\frac{1}{g^2}-\frac{2 t}{g}-t^2-2 g t^3-5 g^2 t^4-14 g^3 t^5
+\CO(g^4).
\ee
\end{example}

\sectiono{Applications to supersymmetric gauge theories}

\subsection{Superpotentials in ${\cal N}=1$ gauge theories}

In \cite{dv,dvp} Dijkgraaf and Vafa argued that superpotentials in a large class of $\CN=1$ supersymmetric gauge theories can be 
computed by using matrix models. Let us consider an $\CN=1$ supersymmetric gauge theory with gauge group $G={\rm U}(N), {\rm SO}(N)$ or ${\rm Sp}(N)$, 
where the superfield strength is denoted by $\CW^{\alpha}$. There is also a chiral 
superfield $\Phi$ in a representation $R$ of the gauge group $G$, with a tree level superpotential $W_{\rm tree}(\Phi)$, which we will assume to be a polynomial of 
degree $d$:
\be
W_{\rm tree}(\Phi)=\sum_{j=1}^d {g_j \over j} \tr \Phi^j.
\ee
If all roots of $W'_{\rm tree}(x)= g_d \prod_{i=1}^{d-1}(x-a_i)$ are distinct,
the matter fields are all massive; a classical vev for $\Phi$, where $N_i$ of
its eigenvalues are equal to $a_i$, spontaneously breaks part of the gauge
symmetry, and the massive fields  can be integrated out to get an
effective action for the unbroken gauge degrees of freedom at low energy. \\

Depending on the gauge group and the 
representation we will end up with different patterns of gauge symmetry breaking (see the useful summary in eq. (2.1) of \cite{shigemori2})). We will be particularly interested in the 
examples where $G={\rm SO}(N), \, {\rm Sp}(N)$ and $R$ is, respectively, the symmetric and the antisymmetric representation of the group. In this case, we have the simple patterns of 
gauge symmetry breaking
\be
\ba
{\rm SO}(N) \rightarrow &\prod_{i=1}^{k}{\rm SO}(N_i), \\
{\rm Sp}(N) \rightarrow &\prod_{i=1}^{k}{\rm Sp}(N_i),
\ea
\ee
where $1\le k \le d-1$, and there will be, correspondingly, various gluino superfields for the unbroken gauge groups, 
\be
S_i =-{1\over 32 \pi} \tr \left( \CW^{(i), \, \alpha} \CW^{(i)}_{ \alpha}\right), 
\ee
where $\CW^{(i)}_{ \alpha}$ is the superfield strength for the $i$-th gauge
group. According to the proposal of \cite{dv, dvp}, the effective
superpotential for the glueball superfields, as well as the gauge coupling
matrix for the infrared-free abelian fields, should be computable from an
auxiliary matrix model with $V(x)=W_{\rm tree}(x)$. In particular, the glueball superpotential, as a function of the gluino superfields, is given by \cite{ino,ks1,shigemori2}
\be
\label{efsuper}
W_{\rm eff}(S_i) =W_{\rm VY}(S_i)+\sum_{i=1}^{k} N_i {\partial F_{0,0} \over \partial S_i}-4 \epsilon  F_{0,1}
\ee
where $\epsilon=\pm1$ for ${\rm SO}/{\rm Sp}$, respectively and $W_{\rm VY}(S_i)$ is the Veneziano--Yankielowicz superpotential 
(see \cite{shigemori2} for a detailed 
expression). In this equation, 
$F_{0,0}$, $F_{0,1}$ are the first two free energies in the expansion (\ref{genusF}), 
obtained in the $\beta$-ensemble for a matrix model with potential
$V(x)=W_{\rm tree}(x)$, in the $k$-cut phase, and with 't Hooft parameters $S_i$. In addition, the gauge theory quantity
\be
T(z) =\left \langle \tr \left( {1\over z-\Phi}\right) \right\rangle
\ee
can be computed from the generalized Konishi anomaly \cite{ac} and expressed in terms of matrix model resolvents \cite{ks2,shigemori2}:
\be
\label{gtres}
T(z)= \sum_{i=1}^k N_i {\partial W_{0,0}(z) \over \partial S_i} -4 \epsilon  W_{0,1}(z),
\ee
where again $\epsilon=\pm1$ for ${\rm SO}/{\rm Sp}$. Similarly, contributions to chiral ring observables induced by a non-flat gravity
background can be computed in terms of non-planar corrections to the resolvent
\cite{ac2}. 

The formulae above for the solution of \eqref{recursive} in the polynomial matrix model
case give then explicit results for computing a large class of vevs of chiral
observables for a general $W_{\rm tree}(x)$. In particular,  $W_{0,1}$ and
$F_{0,1}$ yield the unoriented contribution to the the effective
superpotential (\ref{efsuper}) and gauge theory resolvent (\ref{gtres}) for a
general tree-level superpotential\footnote{In the Appendix A of
  \cite{shigemori2}, the expressions for the unoriented contributions to the
  free energy and the resolvent in terms of planar, oriented contributions,
  are only valid when $k$ (the number of cuts) takes its maximum value $d-1$
  for a given potential.}. As an example and a test of our computations, let
us consider the case of classically unbroken gauge symmetry, where $k=1$. This corresponds to the one-cut case in the computations above. Using the 
well-known one-cut result (see for example \cite{dfgzj})
\be
{\partial W_{0,0}(z) \over \partial z}={1\over {\sqrt{\sigma(z)}}},
\ee
as well as (\ref{simpleonecut}), we find the general formula
\be
T(z)={N \over  {\sqrt{\sigma(z)}}} +{2 \epsilon}{y'(z) \over y(z)} - {2 \epsilon \over {\sqrt{\sigma(z)}}}\left[ d-1+\sum_i {\sqrt{\sigma(z_i)} \over p-z_i}\right].
\ee
This agrees with the explicit computation for the quartic potential in \cite{ac}.

\subsection{Penner model and AGT correspondence}
\label{sec:AGT}
A more sophisticated example is given by the double Penner model
\be
V(x)= \al_1 \log x + \al_2 \log (x-1).
\ee
This model was recently considered by Dijkgraaf and Vafa\footnote{See also
  \cite{Eguchi:2009gf} for further developments.} \cite{dvtoda} in the context of the AGT
correspondence \cite{agt}, where it was shown to give a matrix model
representation  of the chiral three-point function in Liouville theory; its 4d
counterpart arises \cite{Gaiotto:2009we} as the dimensional reduction of the 6d $A_1$ (2,0) theory compactified on a sphere with three punctures, and is a $U(1)$ theory with four hypermultiplets. The spectral curve for this case reads 
\be
M(p)\ssq{p}=\frac{\sqrt{\al_1^2(1-p)+\al_3^2(p^2-p)+\al_2^2 p}}{p(p-1)}
\label{eq:spcurv2p}
\ee
where $\al_3=-t-\al_1-\al_2$ and $t=g_s N$. \\

An extension of the AGT 
correspondence in presence of defects was considered in \cite{aggtv}, where multiple insertions of surface operators on the 4d gauge
theory side were mapped to insertions of vertex operators corresponding to
degenerate states on the Liouville theory side. In \cite{saraetal,dgh} both were mapped in turn to $A$-type
open topological string amplitudes on the toric geometries that engineer the
relevant gauge theory. In particular the authors of \cite{saraetal} conjectured and checked that the Liouville theory four--point function with one degenerate insertion and vanishing background charge $Q=b+1/b$
\be 
Z_{\rm null}(p, \hbar, b=i)=\frac{\bra \frac{\al_1}{\hbar} |  V_{-\al_{2}/\hbar} (1)  V_{-b/2}(p)| -\frac{\al_3}{\hbar}+\frac{b}{2}  \ket}{ \bra \frac{\al_1}{\hbar} |  V_{-\al_2/\hbar} (1) | -\frac{\al_3}{\hbar}  \ket}\Bigg|_{b=\ri}   \,,
\label{eq:znull}
\ee
should be expressible in terms of oriented topological string amplitudes computed through the Eynard-Orantin recursion applied to \eqref{eq:spcurv2p}
\be 
Z_{\rm null}(p, \hbar, b=i)= \exp\l[\frac{1}{\hbar}A_1^{(0)}(p)+\frac{1}{2!}A_{2}^{(0)}(p,p)+\hbar \l(A_1^{(1)}(p)+\frac{1}{3!}A^{(0)}_3(p,p,p) \r) +\dots\r],
\label{eq:4pfor}
\ee
with
\be
A_{h}^{(g)}(p_1, \dots, p_h)=\int \rd p_1 \dots \rd p_h W_{h}^{(g)}(p_1, \dots, p_h).
\ee
On the other hand, it was proposed in \cite{dvtoda} that turning on a background charge $Q$ on the CFT side should exactly correspond to the $\beta$-deformation of the matrix model, with the dictionary been given by
\be
Q^2=
-\gamma^2, \quad b^2=-\beta,  \quad \hbar=\frac{g_s}{\sqrt{\beta}}.
\label{eq:dictqb}
\ee
This was checked by direct computation in \cite{schiappaw,morozov, morozov2, Itoyama:2010ki} at the level of the free energy. 
It is therefore tempting to look at a combination of the two claims above and compute refined open string amplitudes via \eqref{recursive}, corresponding to degenerate insertions in Liouville theory with non-vanishing $Q$.
A natural extension of \eqref{eq:4pfor} in the $\beta$-deformed case is through an expansion of the form 
\be
\ba
F_{\rm null}(p,\hbar,b)&=\log Z_{\rm null}(p,\hbar, b)\\
&=\sum_{n=-1}  \left(\sqrt{\beta}\hbar\right)^{n} \sum_{g,h,k|2g-2+h+k=n} \frac{1}{h!} \beta^{1-g-k/2} A^{(g)}_{h,k}(p,\cdots p) \gamma^k.
\ea
\label{eq:4pfun}
\ee
The $\beta$-deformed topological recursion allows us to test this proposal in detail. On the CFT side it is well-known that Ward identities for the normalized four point function \eqref{eq:znull} reduce to a hypergeometric differential equation; more precisely we have that
\be 
Z_{\rm null}(p,\hbar,b)=p^{\frac{
b \al_1}{\hbar}}(1-p)^{\frac{-b \al_2}{\hbar}} \,{}_2 F_1(A_1,A_2;B_1;p)\,,
\ee
where 
\be A_1 = b\, \frac{\al_3-\al_2+\al_1}{\hbar}\,,\qquad  A_2=b\, \l(\frac{\al_1-\al_2-\al_3}{\hbar} + Q\r)\,, \qquad B_1= \frac{2 b \al_1}{\hbar}  \,.
\ee
By Taylor expanding around $\hbar=0$, $b=i$ we obtain
\ben
\log Z_{\rm null}(p) &=&
\frac{b}{\hbar}A^{(0)}_{1,0}(p)+\l[ -\frac{b^2}{2}A^{(0)}_{2,0}(p,p)+
(b^2+1)A^{(0)}_{1,1}(p)\Big)\r] \nn \\
&+& 
\frac{\hbar}{2 b}\l[(1+b^2)^2 A^{(0)}_{1,2}(p)-b^2(b^2+1)\frac{1}{2} A^{(0)}_{2,1}(p,p)-b^2A^{(1)}_{1,0}(p)+b^4\frac{1}{3!}
A^{(0)}_{3,0}(p,p,p)
\r]\nn \\
&+& \frac{\hbar^2}{4 b^2} \Bigg[ (b^2+1)^3 A^{(0)}_{1,3}(p)-b^2(1+b^2)A^{(1)}_{1,1}(p)+b^4\frac{1}{2}A^{(1)}_{2,0}(p,p)-(1+b^2)\nn \\
& & b^2 \frac{1}{2} A^{(0)}_{2,2}(p,p)
-b^6\frac{1}{4!} A^{(0)}_{4,0}(p,p,p,p)+\frac{1}{3!}b^4 (b^2+1)A^{(0)}_{3,1}(p,p,p)
\Bigg]+\mathcal{O}\left(\frac{\hbar}{b}\right)^3. \nn \\
\label{eq:cftexp}
\een

On the other hand, we can apply the refined recursion to the spectral curve \eqref{eq:spcurv2p}. In this case, the contour integrals also have contributions from the poles of $M(x)$; as an instance, we find for the one-crosscap correction to the resolvent
\be
W_{0,1}(p)=-{1\over 2} {y'(p) \over y(p)} +{1\over 2 {\sqrt{\sigma(p)}}}   \sum_{i=1}^3 {\rm Res}_{z=z_i} \left[{1\over p-z} \left( {M'(z) \over M(z)} \sqrt{\sigma(z)} + {1\over 2} {2z-a-b \over  {\sqrt{\sigma(z)}}}\right) \right],
\ee
where
\be
z_1=0,\quad z_2=1, \quad z_3=\infty. 
\ee
The residues give the values
\be
-{\alpha_1 \over \alpha_0}{1\over p} , \qquad -{\alpha_2 \over \alpha_0}{1\over p-1}, \qquad -1, 
\ee
for $i=1,2,3$, respectively, and we finally obtain
\be
W_{0,1}(p)=-{1\over 2} {M'(p) \over M(p)} -{1\over 4} {2p-a-b \over (p-a)(p-b)} -{1\over 2 {\sqrt{\sigma(p)}}} \left( 1+ {\alpha_1 \over \alpha_0}{1\over p}+ {\alpha_2 \over \alpha_0}{1\over p-1}.
\right). 
\ee
The integrated refined amplitudes $A^{(g)}_{h,k}(p,\cdots p)$ can be similarly computed in a straightforward fashion from \eqref{recursive}; upon taking into account the dictionary \eqref{eq:dictqb}, we find exact agreement with the CFT expansion \eqref{eq:cftexp}.

\sectiono{The $\beta$-deformed Chern--Simons matrix model}

\subsection{Definition and relation to the Stieltjes--Wigert ensemble}
The $\beta$-deformed Chern--Simons (CS) matrix model on $\IS^3$ is defined by the partition function 
\be
\label{csbeta}
Z_{\rm CS}(N, g_s, \beta)={1\over N!} \int \prod_{i=1}^N {\rd x_i \over 2\pi} \re^{-{\beta \over 2 g_s} \sum_{i=1}^n x_i^2} \prod_{i<j} \left(  2 \sinh {x_i-x_j \over 2} \right)^{2\beta}.
\ee
When $\beta=1$ we recover the standard CS matrix model considered in \cite{mm,mmhouches}. This generalization of the CS matrix model is the natural counterpart of the 
$\beta$-ensemble deformation of the standard Hermitian matrix model. \\

In \cite{tierz} Tierz pointed out that the standard CS matrix model could be written in the usual, Hermitian form, i.e. with a Vandermonde inteaction among eigenvalues, 
but with a potential 
\be
\label{logpot}
V(x)={1\over 2} \left( \log x\right)^2. 
\ee
This potential defines the so-called Stieltjes--Wigert (SW) matrix model. 
It is very easy to show that (\ref{csbeta}) is, up to a simple multiplicative factor, the partition function of the $\beta$-deformed version of the SW matrix model. 
To do that, we perform the change of variables
\be
\label{changev}
u_i=c \,  \re^{x_i}
\ee
where $c$ is given by
\be
c=\exp \left( t-g_s (1-\beta^{-1}) \right),  \qquad t=g_s N. 
\ee
A simple computation shows that 
\be
Z_{\rm CS}(N, g_s, \beta)= \re^{-{\beta N \over 2g_s} (\log c)^2} Z_{\rm SW}(N, g_s, \beta) 
\ee
where
\be
Z_{\rm SW}(N, g_s, \beta) ={1\over N!} \int \prod_{i=1}^N {\rd u_i \over 2\pi} \re^{-{\beta \over 2 g_s} \sum_{i=1}^n (\log u_i)^2} \prod_{i<j} \left(u_i-u_j \right)^{2\beta}
\ee
is the partition function of the $\beta$-deformed SW ensemble. In terms of free energies we have
\be
\label{eq:SW2CS}
F_{\rm CS}(N, g_s, \beta)=-{\beta t^3  \over 2}g_s^{-2} + g_s^{-1} (\beta-1) t^2 -(\beta+\beta^{-1}-2) {t\over 2} +F_{\rm SW}(N, \beta, g_s).
\ee
The change of variables (\ref{changev}) has to be taken into account when computing correlation functions in the CS matrix model from the SW matrix model, and we have 
the relationship
\be
\label{changecor}
\langle \tr \, U^n \rangle^{\rm SW}=\exp\left( n t -n g_s (1-\beta^{-1}) \right) \langle \tr \, \re^{n X} \rangle^{\rm CS}, 
\ee
where 
\be
U={\rm diag}(u_1, \cdots, u_N), \quad X={\rm diag}(x_1, \cdots, x_N).
\ee

It was shown in \cite{mmhouches} that, though both the potential (\ref{logpot}) and its first derivative are non-polyomial, 
the SW model can be solved at large $N$ with standard saddle-point techniques. In particular, the resolvent is given by 
\be
W_0(p)=-{1\over p}\log\left[ {1+\re^{-t}p +{\sqrt{(1+\re^{-t}p)^2-4 p}} \over 2p}\right], 
\ee
and the spectral curve is 
\be
y(p)=M(p) {\sqrt{ (p-a)(p-b)}}={2\over p}\tanh^{-1} \left[ {{\sqrt{(1+\re^{-t}p)^2-4 p}} \over  1+\re^{-t}p}\right] , 
\ee
where 
\be
M(p)={2\over p  {\sqrt{ (p-a)(p-b)}}} \tanh^{-1} \left[ {{\sqrt{(1+\re^{-t}p)^2-4 p}} \over  1+\re^{-t}p}\right]
\ee
and the positions of the endpoints are given by 
\begin{eqnarray}\label{endpoints}
a(t)&=&2\re^{2t}-\re^t +2\re^{3 t \over 2}{\sqrt {\re^t-1}}, \nonumber\\
b(t)&=&2\re^{2t}-\re^t -2\re^{3 t \over 2}{\sqrt {\re^t-1}}.
\end{eqnarray}
For $t=0$, $a(0)=b(0)=1$, which is indeed the minimum of (\ref{logpot}).

\subsection{Corrections to the resolvent and to the free energy}

The SW ensemble is, from many points of view, a conventional one-cut matrix model, and its correlation functions and free energies obey the standard recursion relations of \cite{eo,ce}. We now proceed to calculate the first $\beta$-deformed corrections to the resolvent and the free energy by using the recursion of \cite{ce}. 

Let us first consider the correction fo the 1-point correlator (\ref{w01}). As in the standard polynomial case, there is no contribution from $V''(q)$ (since both this function and $M(q)$ are analytic on the cut). In order to proceed, it will 
be useful to change variables from $q$ to $\zeta$ through,
\be
\label{changezeta}
q={b-a\over 2} \zeta +{a+b\over 2}.
 \ee
 This maps the interval $[a,b]$ to $[-1, 1]$. Explicitly, 
 \be
 q=2 \re^t  \sqrt{\re^t(\re^t-1)}  \zeta+\re^t \left( 2 \re^t-1\right).
 \ee
 In terms of $\zeta$ we have
 \be
 \tanh^{-1} \left[ {{\sqrt{(1+\re^{-t}q)^2-4 q}} \over  1+\re^{-t}q}\right] =  \tanh^{-1} \left( {{\sqrt{\zeta^2-1}} \over \zeta+ (1-\re^{-t})^{-1/2}} \right),
 \ee
 and the moment function reads 
 \be
M(\zeta)={1 \over \left( c_1 +c_2 \zeta \right) \sqrt{\zeta^2-1}} \tanh^{-1} \left( {{\sqrt{\zeta^2-1}} \over \zeta+ (1-\re^{-t})^{-1/2}} \right), 
\ee
where
\be
\label{mcoeffs}
c_1= \re^{5t/2}\sqrt{\re^t-1} \left(2 \re^t-1 \right), \quad c_2 = 2 \re^{3 t} \left(\re^t-1\right).
\ee
The integrand of (\ref{w01}) involves then, 
\be
\label{intcrazy}
-{1\over 2} \rd p {y'(q) \over M(q)} ={1\over 2 q} {\sqrt{(q-a)(q-b)}} \rd q - \re^{ t}  \CM(\zeta)  \rd \zeta, 
\ee
with 
\be
\CM(\zeta)={  {\sqrt{\re^{ t} (\re^t-1)}}  \left( \sqrt{ \re^t( \re^t-1)}  \zeta+\re^t-1 \right) \over   \left(2 \sqrt{\re^t(\re^t-1)}  \zeta+2 \re^t-1\right)  \tanh^{-1} \left( {{\sqrt{\zeta^2-1}} \over \zeta+ (1-\re^{-t})^{-1/2}} \right)}.
\ee
We then obtain, 
\be
\label{crazyres}
W_{0,1}(p)={1\over 2} {p-\re^t \over p {\sqrt{\sigma(p)}}}-{1\over 2p} -{ \re^t \over  {\sqrt{\sigma(p)}}} \oint_{\CC} {\CM(\zeta) \over p-2 \re^t  \sqrt{\re^t(\re^t-1)}  \zeta-\re^t \left( 2 \re^t-1\right)} {\rd \zeta \over 2 \pi \ri},
\ee
where the integral involving the first term in (\ref{intcrazy}) has been calculated through a contour deformation and picking residues at $q=0, \infty, p$, and the contour $\CC$ encircles the cut 
$[-1,1]$ in the $\zeta$ variable. 

We have not been able to calculate the second term in (\ref{crazyres}) in closed form. In order to obtain explicit results, we have to perform a series expansion in both $p$ and $t$. To see 
an explicit example of this procedure, we expand around $p=\infty$ to obtain
\be
\label{wsq}
W_{0,1}(p) \Big|_{p^{-2}}=-\re^{t}(\re^t-1) + \re^t S(t) 
\ee
where
\be
\label{intS}
S(t)= {1\over  \pi \ri}\int_{-1}^1 \rd \zeta  \, \CM(\zeta).
\ee
Notice that this integral depends on $t$ only through the variable $\nu=\re^t$. It can be computed systematically as a power series in $\nu-1$, which can then be re-expanded 
as a power series in $t$. We obtain, for the first few orders,
 \be
 S(t)= -\frac{t}{2}-\frac{7 t^2}{24}-\frac{71 t^3}{720}-\frac{2971 t^4}{120960}-\frac{17809 t^5}{3628800}-\frac{4843
   t^6}{5913600}-\frac{51012187 t^7}{435891456000}+\CO\left(t^8\right).
\label{eq:st}
 \ee
The planar limit of the vev of $\tr\, U$ is given by
 \be
  \langle \tr \, U \rangle^{\rm SW}_{0,0}=\re^t(\re^t-1), 
  \ee
and its first correction is given by (\ref{wsq}), 
\be
\langle \tr U \rangle^{\rm SW}_{0,1} = W_{0,1}(p) \bigg|_{p^{-2}}= \re^t \left( S(t) -\re^t +1 \right).
\ee
Together with (\ref{changecor}) we then deduce that 
\be
  \langle \tr \, \re^x\rangle^{\rm CS}_{0,1}= S(t). 
\label{eq:trmcs}
  \ee
 On the other hand, a direct perturbative computation of the vev $\langle \tr \, \re^x\rangle^{\rm CS}$ in the CS matrix model (\ref{csbeta}) gives
\be
\ba
\langle \tr \, \re^x\rangle^{\rm CS}&=N + g_s \left[ {N^2 \over 2} -{1\over 2} (1-\beta^{-1}) N\right]
\\ 
&+ g_s^2 \left[ {N^3 \over 6} -{7\over 24}  (1-\beta^{-1}) N^2 - {1\over 24} \beta^{-1} N  +{1\over 8} (1-\beta^{-1})^2 N \right] \\
&+ g_s^3\biggl[ {N^4 \over 24} -{71 \over 720} N^3(1-\beta^{-1}) -{1\over 48} \beta^{-1} N^2 +{7 \over 90}(1-\beta^{-1})^2 N^2 \\ 
& -{1\over 48} (1-\beta^{-1})^3 N +{11\over 720} \beta^{-2} (\beta-1)N \biggr] +\CO(g_s^4),
\ea
\ee
in complete agreement with (\ref{eq:trmcs}).

Using the same type of techniques we can also compute the first correction to the free energy. Using (\ref{derexf}) we find, 
\be
\partial_t F^{\rm SW}_{0,1} = 1 +  \int_a^b{\rd p  \over  \pi} {\log|M(p)| \over
  {\sqrt{|\sigma(p)|}}} + \int_a^b{\rd p \over 2 \pi} {\log  |\sigma(p)| \over
  {\sqrt{|\sigma(p)|}}}.
\label{integrals}
\ee
The last integral can be computed exactly:
\be
 \int_a^b{\rd p \over 4 \pi} {\log  |\sigma(p)| \over
  {\sqrt{|\sigma(p)|}}}= \frac{\log(1-\re^{-t})}{2}+2 t.
\ee
The first integral can be written, using again the change of variables (\ref{changezeta}), as 
\be
\ba
\int_a^b{\rd p  \over  \pi} {\log|M(p)| \over
  {\sqrt{|\sigma(p)|}}} =
\int_{-1}^1{\rd\zeta \over \pi} \frac{1}{ \sqrt{1-\zeta^2}}
\log \left(\frac{\tan ^{-1}\left(\frac{\sqrt{1-\zeta^2}}{\zeta+\frac{1}{\sqrt{1-e^{-t}}}}\right)}{
\left(c_1 + c_2 \zeta\right) 
\sqrt{1-\zeta^2}}\right),
\ea
\ee
where $c_1$, $c_2$ are defined in (\ref{mcoeffs}). 
As before, this integral can be computed as a power series in $t$ around $t=0$. Putting everything together, and taking into account \eqref{eq:SW2CS},  we find
\be
\label{csb}
F^{\rm CS}_{0,1}(t)=
\frac{1}{2} (\log (t)+1) t-\frac{t^2}{12}+\frac{t^3}{1440}+\frac{17 t^4}{45360}-\frac{137 t^5}{14515200}-\frac{2 t^6}{467775}+\CO(t^7).
\ee
We have again verified the very first coefficients in this expansion against a direct perturbative calculation in the CS matrix model. \\

 It is worth pointing out that the corrections appearing in the CS matrix model when $\beta\not=1$ are much more complicated than the ``standard" ones. 
For example, for $\beta=1$ all the correlators are polynomials in $\re^t$, while the integral giving (\ref{intS}) is not. 

\subsection{$\beta$-deformation and the $\Omega$ background}
One of the interesting aspects of the conventional CS matrix model with $\beta=1$ is that its large $N$ expansion equals the $1/N$ expansion of topological string 
theory on the resolved conifold \cite{gv}, since it equals the partition function of CS theory on the three-sphere. On the other hand, the partition function 
of topological string theory on the resolved conifold admits a refinement given by the K-theoretic version of Nekrasov's partition function for a $U(1)$ theory \cite{no}. This 
partition function can be also obtained from the refined topological vertex of \cite{ikv}. The first correction to the refined free energy of the resolved conifold is simply 
\be
F_{0,1}^{\rm ref}={1\over 2} {\rm Li}_2(\re^{-t}).
\ee
This expression is much simpler than the result (\ref{csb}). 
\\

 We can also compare the result for $\langle \tr \, \re^x \rangle^{\rm CS}$ with expectations coming from the theory of the refined vertex. When 
$\beta=1$, the correlation function $\langle \tr \, \re^x \rangle^{\rm CS}$ can be expressed in terms of the open string amplitude 
\be
 Z_{\tableau{1}}(t,g_s)= {1 -Q \over 2 \sinh\left({g_s\over 2}\right)}. \nn
\ee
for a D-brane in an external leg of the 
resolved conifold. Here, $Q=\re^{-t}$. The precise relation involves a framing factor, 
\be
\label{openrel}
\langle \tr\,  \re^x \rangle^{\rm CS}_{\beta=1}=\re^{t} Z_{\tableau{1}}(t,g_s).
\ee
The ``refined'' version of the D-brane amplitude is\footnote{This form of the amplitude is dictated by an implicit choice
  of gluing along one of the unpreferred legs of the refined topological vertex. Other choices of
  gluing only result in minor differences in this particular case, which for the $\mathcal{O}(1)$ term
  in \eqref{refzexp} amount to an overall rescaling by a factor of $\re^{t}$;
  this obviously leaves unchanged the discussion about the comparison with the
  Chern-Simons matrix model computation.} 
\be
\label{refz}
Z_{\tableau{1}}(t,g_s, \beta)= \frac{Q \sqrt{q_2}-\sqrt{q_1}}{q_2-1}
\ee
with
\be
q=\re^{-\sqrt{\beta} g_s}, \quad t=\re^{-g_s/\sqrt{\beta}}.
\ee
Expanding in $g_s$ and $\beta$ we obtain 
\be
Z_{\tableau{1}}(t,g_s, \beta)= \l(1-\re^{-t}\r)\frac{1}{g_s} + \frac{1}{2\sqrt{\beta}}(1-\beta) +\cdots, 
\label{refzexp}
\ee
so it is clear that the relationship (\ref{openrel}) is no longer true when we consider the $\beta$-deformed CS ensemble in the l.h.s., and the refined amplitude (\ref{refz}) in the r.h.s. \\

Of course, in the comparisons we have made, we assumed that the 't Hooft parameter $t_{\rm CS}$ in the matrix model 
is equal to the parameter $t_{\rm TS}$ appearing in the refined topological string. We have not excluded the possibility that both sides are related by a more general relation of the form, 
\be
\label{relf}
t_{\rm CS}=t_{\rm TS} +f(\beta, t_{\rm TS})
\ee
where $f (\beta, t_{\rm TS})$ vanishes for $\beta=1$ (since the two parameters agree in that case). But in order to reproduce the above results, the unknown function in (\ref{relf}) should be 
rather complicated. Another possibility is that we have to modify as well the Gaussian potential in order to match the $\Omega$-deformed topological string. This has been 
suggested in a closely related context in \cite{sulkowski}. 

\vskip 1truecm

\noindent  {\large{\bf Acknowledgements}}\\

 We are especially grateful to S.~Pasquetti for many useful discussions and
collaboration at an early stage of this project. We would also like to thank 
M.~Aganagic, B.~Eynard, H.~Fuji, A.~Klemm, C.~Koz\c caz, D.~Krefl, N.~Orantin, C.~Vafa and N.~Wyllard for
discussions and/or email correspondence. This work was partially supported by
the Fonds National Suisse (FNS).

\appendix
\section{Useful formulae for elliptic integrals}
In this section we collect a few formulae regarding the
expansion of elliptic integrals for small values of the elliptic modulus, which are relevant
for the computations of Section \ref{sec:twocut}.
\ben
\label{eq:exppi}
\Pi (n|m) &=& \frac{\pi}{2} \sum_{k=0}^\infty\frac{m^k}{(k!)^2}\left(\frac{1}{2}\right)_k^2 \left(\frac{n^{-k} k!}{\sqrt{1-n} \left(\frac{1}{2}\right)_k}-\frac{2 k
   \sum_{j=0}^{k-1}\left(\frac{\left(1-\frac{1}{n}\right)^j (1-k)_j}{\left(\frac{3}{2}\right)_j}\r)}{n}\right) \\
K(m) &=& \frac{\pi}{2} \sum_{k=0}^\infty \frac{m^k \left(\frac{1}{2}\right)_k{}^2}{(k!)^2}
\label{eq:expk}
\een

\end{document}